# Tokamak operation at low q and scaling toward a fusion machine

R. Paccagnella^

Consorzio RFX, Associazione Euratom-ENEA sulla Fusione, Padova, Italy

^ and Istituto Gas Ionizzati del Consiglio Nazionale delle Ricerche (CNR), Padova, Italy

## Abstract:

Operation of a tokamak with q edge around 2 is discussed in this paper.
It is shown that an L mode operation at a relatively high toroidal magnetic field can produce confinement conditions similar to ITER-FEAT.

## Introduction:

The research to make fusion power achievable to the mankind has mainly dedicated his efforts, in the last 40-50 years, to develop the tokamak concept.
The result is the big international collaboration effort which aims to build the ITER fusion device in the next years.
Although ITER will be a huge step toward fusion, it is a very complicated and costly machine. Moreover it is unclear at the moment if some basic drawbacks in ITER (and in general in the tokamak concept has it has been evolved) can really be overcome to build a future commercial reactor along the same conceptual lines.
In particular the three major obstacles to scale ITER-like machines to a reactor are:
a) the plasma disruptions
b) the presence of localized modes at the plasma edge (ELMs)
c) the high localized power exhaust need at the divertor plates
The points (a) and (b) are both connected with the presence of plasma instabilities which are very difficult to foreseen and to control. While point (c) is more related to the way in which the tokamak concept has evolved in the last 20 years. The presence of divertors plates are connected with the necessity of producing a magnetic topology with open field lines at the boundary between the plasma and the wall. This has the advantage of controlling the plasma wall interaction and high impurities influxes from the wall. However the drawback is an extremely high power density deposited at the divertor plates, which in ITER will be at the limit of sustainability for the plates material, while in a reactor will be well above the threshold for fusion (thermal this time!) for any known material.

In this paper an attempt is made to overcome these difficulties by proposing a different approach. In particular it is proposed to overcome (a) and (b) by operating a tokamak in a region of parameters which was considered previously as forbidden, i.e. a low q (safety factor) operation. This, as will be discussed below, could be done with an active control system.

A solution to point (c) instead is based simply upon the removal of the divertors by operating at high plasma density with a low Z material wall. In this way the plasma-wall interaction is spread over the entire wall surface lowering the power per unit area to reasonable (1 $MW/m^2$) levels.

**A low q tokamak at high density:**

Recently low q controlled ohmic discharges has been obtained in RFX-mod [1].

The detailed discussion of the control system is outside the scope of this paper, however the main feature that make the control system in RFX able to stabilize the low q tokamak is the presence of several (192) active coils each equipped with is own power supply. This permit to generate a very clean spectrum of Fourier harmonics and the avoidance of dangerous sidebands which can couple different magneto-hydro-dynamic ( MHD ) modes. A recent detailed discussion of this issue has been presented at the $16^{th}$ MHD stability and control workshop in San Diego [2].

To understand the peculiarity of the achieved parameters the typical operational point can be put in the Hugill diagram (see below). The cross represents roughly the new operational region explored by RFX-mod.

The Hugill diagram is reproduced (for Reader's convenience) from Reference [3].

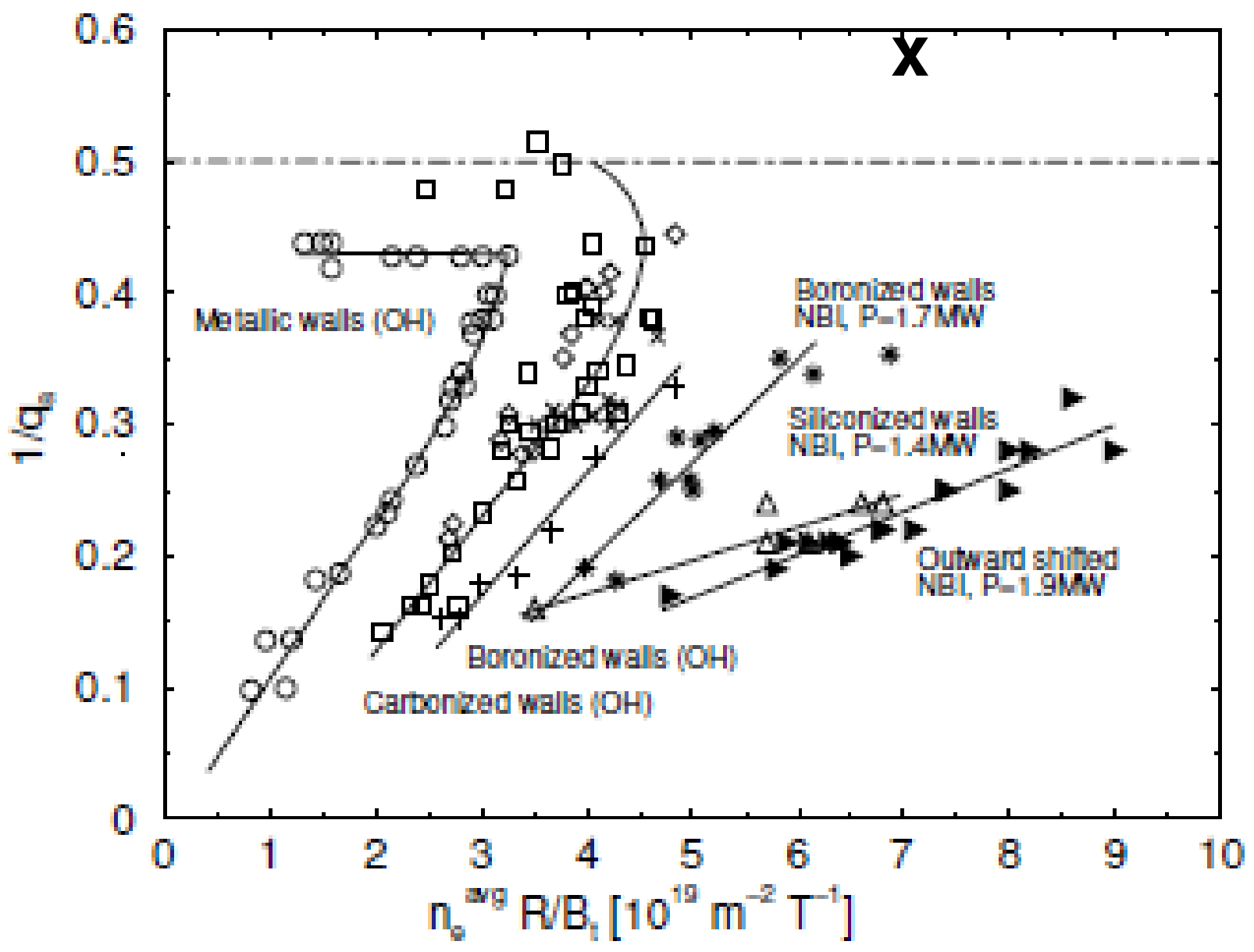

It can be seen immediately that the new point corresponds to a q(a) less than 2 (above the dashed horizontal line) and high operational density.

As regard the confinement the data agree very well with the L mode scaling [4]:

$$\tau_{th} = 0.016\, I^{0.84}\, B^{0.12}\, R^{1.55}\, n^{0.46}\, a^{-0.83}\, A^{0.63}\, M^{0.21}\, P^{-0.76} \quad (1)$$

where I, B, R, a, n have the obvious meaning of plasma current (MA), toroidal magnetic field (T), plasma major and minor radii (m), plasma density ($10^{19}$ $m^{-3}$) respectively, while A ($m^2$ ) is the cross section area, M is the isotopic mass and P (MW) the total input power. It should also be noted that RFX-mod is operating in ohmic regime, since no auxiliary heating systems are available at the moment. Even without auxiliary heating the RFX experimental confinement follows quite well the predictions based on Eq.(1) when as the input power (P) the ohmic power is used.

**Projection to a fusion regime:**

Given the encouraging stable operation around q(a) =2 it is worthwhile to see if these results can favourably scales toward a fusion reactor operating in L mode.
We can start assuming that the density, as it is confirmed by our experiments, will obey the above Hugill's scaling. Therefore we can deduce the density once that we have fixed the major radius and the toroidal magnetic field. If we assume to have a machine similar to ITER but operating at a smaller q(a), we can start assuming around the same ITER major radius, R=6 m. We don't want the aspect ratio to be too tight, for stability and control reasons, therefore let assume a=2 m, which corresponds to R/a=3.

Assume then to operate the machine at the same ITER maximum current level, 15 MA. Since we are speculating about a simpler machine (possibly with a circular cross section) and q(a) =2, we can now calculate the needed toroidal field, which is of 9 T for the maximum current level. Therefore we end up with a toroidal field which is around twice the ITER value. Remember however that the design is much simpler, just a circular machine with internal walls covered by carbon (or in any case a low Z material). Knowing the major radius and the toroidal magnetic field we can extract the density level from the Hugill diagram. For q(a)=2 we can deduce a density with carbon walls of around 8 $10^{19}$ $m^{-3}$ .

To calculate the confinement time we have to know the power needed to heat the plasma to a thermonuclear level (at around 10 KeV). This is clearly a very uncertain point. However let assume that we need something in the order of 100 MW (which is twice the beam power projected for ITER). Assuming further M=2.5, we can calculate the confinement time from the L scaling law to be around 1 second (0.86 s). This will correspond for the given electron density to be around the Lawson limit i.e. around 0.7 $10^{20}$ ($m^{-3}$ s ) and therefore a triple product of the order of 7-8 $10^{20}$ , i.e almost 1 order of magnitude less than that foreseen for ITER.

Note also that the power density per unit area going to the wall is quite modest, around 0.2 $MW/m^2$ .

It could surprise that at q around 2 the L-mode scaling predicts not so unfavourable conditions for fusion (although still below the ITER triple product value), since in Eq.(1) there is no explicit q dependence. However as already reported, for example in [5], the L-mode scaling expressed in physical dimensional quantities by Eq.(1) , is compatible with a dimensionless scaling in terms of the safety factor of the type, $q^{-\alpha}$, with $\alpha$ of order 1. Even more pessimistic scaling of the confinement on q, like $q^{-2}$ , are predicted on the basis of resistive ballooning derived scaling laws [6]. As a consequence it seems an advantage to operate at the lowest possible value of q(a).

Another point that can be a source of some surprise is the fact that a toroidal field of 9 T is needed for a circular tokamak otherwise similar to ITER, which has a toroidal magnetic field of only 5.3 T. This is the consequence of having a circular plasma. In fact, it is known, that for an elongated plasma the toroidal field can be expressed as:

$$B = \frac{2\mu_o I\, R}{2\pi a^2} \; \frac{q_a}{(1+e^2)}$$

where e is the elongation, I is the toroidal current. It is clear that for the same $q_a$ , a, R and I, just the geometrical effect of elongation allows to decrease considerably the toroidal field. For example in the ITER-like machine with an elongation of about 1.8 , having all the other parameters fixed to the values given above, the necessary toroidal field will decreases to 4.2 T. From Ref. [7] it can also be seen that the energy confinement scaling can be explicitly written in terms of the plasma elongation :

$$\tau_{th} = 0.023\, I^{0.96}\, B^{0.03}\, R^{1.83}\, n^{0.4}\, (\frac{R}{a})^{0.06}\, e^{0.64}\, M^{0.2}\, P^{-0.73} \qquad (2)$$

If we assume e=1.8 and B= 4.2 T, with all the remaining parameters as above (but the density rescaled to the new toroidal field becoming 3.5 $10^{19}$ $m^{-3}$), we end up again with a confinement time of around 0.9 second but at a much lower magnetic field. Therefore it seems that the main advantage of an elongated configuration is the strong reduction of the required toroidal field, at the expenses however of a lower operating density and therefore triple product. It should also be noted that the confinement time estimate from Eq.(1) and (2) are not in complete agreement (not surprisingly since these are, after all, empirical laws). A plot summarizing the differences of the confinement times deduced from Eq.(1 ) and Eq.(2) as a function of the power for the just discussed elongated case is shown in Fig.2. It can be seen that Eq.(2) gives somehow more optimistic predictions.

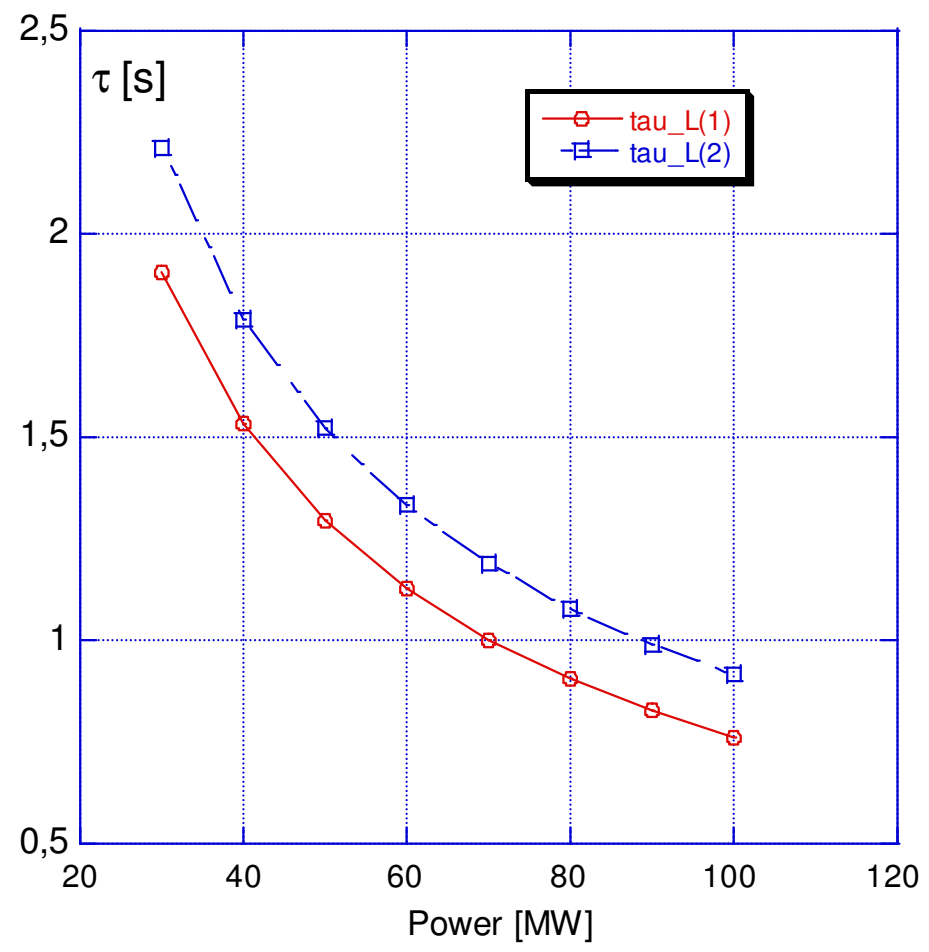

***Fig.2:*** *Confinement time as deduced from eq.(1) and eq.(2) for e=1.8, a=2 m, R=6 m and B=4.2 T as a function of the total power.*

From the scaling of density with the major radius, it can be seen that configurations having a smaller major radius, can allow to operate at higher density. Therefore we analyze the case with R=3, a=1, e=1.8. This correspond to a B=8.4 T at 15 MA and a density exceeding 1.4 $10^{20}$. Plugging these numbers in Eq.(2) a confinement time of about 0.8 seconds is found again, if an input power of around 50 MW (which is higher that the plasma volume assumed reduction) is considered and therefore a triple product which approaches the ITER value (being smaller by a factor 3-4).

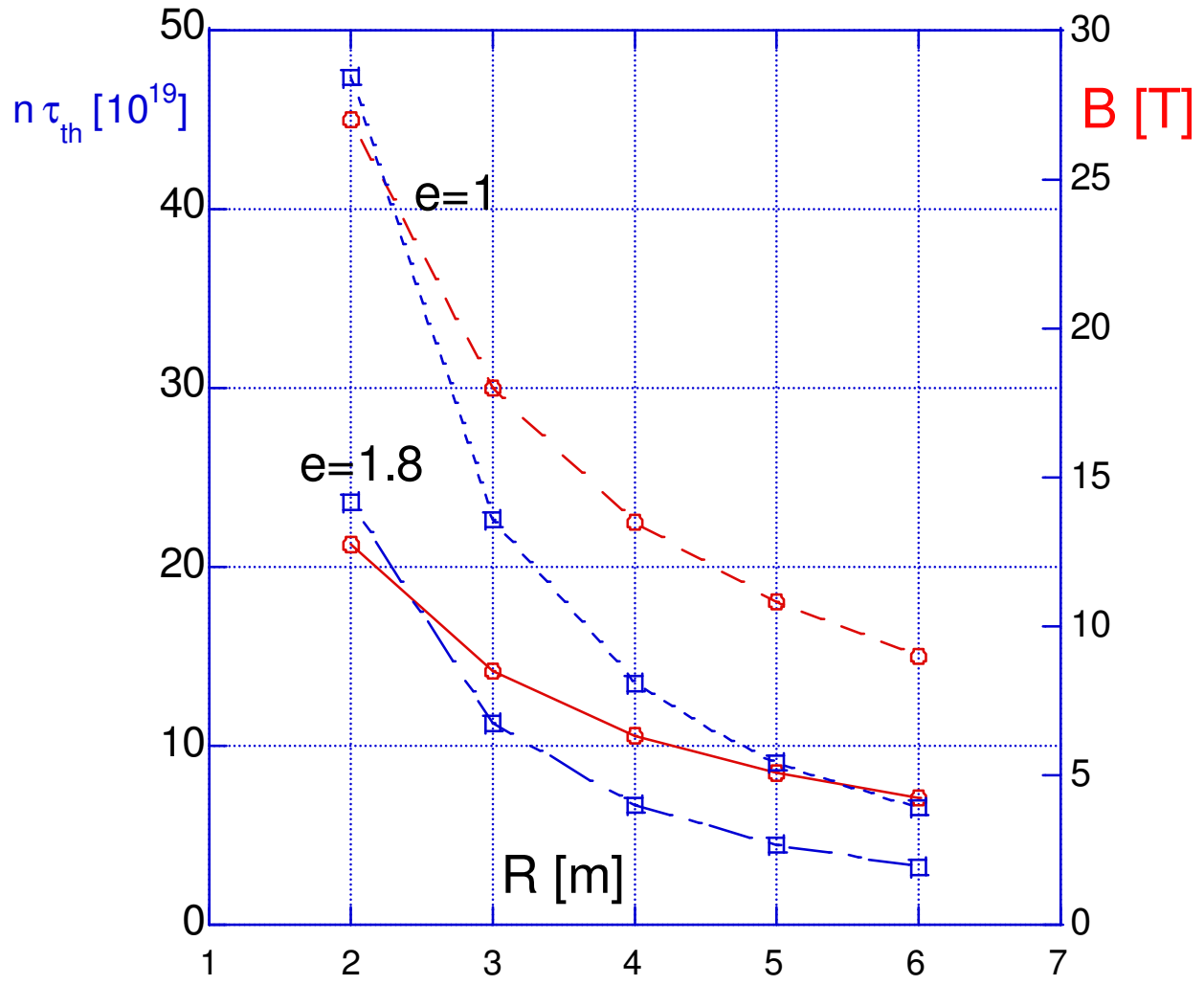

***Fig.3*** *: Lawson parameter (blu squares) and toroidal magnetic field (red triangles) vs. torus major radius (with I=15 MA, e=1.8 and 1, R/a=3, M=2.5 and $n_G$ (Greenwald density) =5 )*

In Fig.3 the results are summarized for different torus radii in terms of the Lawson's product (n $\tau_{th}$ ) and the required toroidal magnetic field. It has been assumed that the input power, P, goes linearly from a value of about 30 MW at R=2 to a value of 100 MW at R=6, somehow consistently (but not exactly proportional) with the plasma volume. It is clear that to obtain high Lawson parameters the machine should be compact (small R) and have an high magnetic field, to increase the density. It is also evident from the figure that for a circular plasma (with aspect ratio 3) the major radius should be higher than 4 m to allow for reasonable and achievable magnetic field strengths. It should also be noted that the question about how easy is the control of the 2/1 mode for an elongated plasma, should be experimentally tested in order to demonstrate the practical possibility of operating in this way. It is in fact a priori clear that the elongation can couple more strongly toroidal sidebands and therefore can in principle complicate the "clean" control of the 2/1 mode [1]. On the other hand it can be easily seen that the case with ellipticity 1.8 has a Troyon $\beta$ = 2.8 I /(a B) ([MA/m T] ) around 5% which can be considered almost reasonable for a reactor (while the e=1 case has $\beta$ around 2% ).

From the above considerations and discussion it is clear that our results point to a similar philosophy as the Ignitor project [8], since high density and high toroidal field seem to give clear advantages. The main difference being that an higher current can be assumed (for a given toroidal field) since the q(a) is lower here. This is not a negligible difference, since the ohmic power is expected to be the main heating source (both in Ignitor and in the present concept). Moreover the lower q at the edge and the operation in L mode can hopefully help against long wavelength magneto-hydro-dynamic (MHD) modes (for example the m=3/n=2 mode) and certainly eliminate edge localized modes.

Similarly to Ignitor the problem of additional heating systems remain quite open, due to the high plasma density and high magnetic field. Note however that the operational density that we have considered here is about ½ of the one assumed for Ignitor, which is a consequence of having fixed the density to the value that can be deduced from experimental data with a carbon wall. Moreover at low q extra ohmic power is available for heating therefore it cannot be excluded that the proposed device could possibly ignite without the necessity of additional non-ohmic power.

Another important point that should also be taken seriously into account is related to the field errors (i.e. unavoidable magnetic field misalignment in a real device). It is well known that field errors have a detrimental effect on magnetically confined plasmas. One clear experimental result, still not fully understood, is that field errors are shielded more efficiently at high plasma density although also a weaker detrimental scaling with the applied toroidal magnetic field is generally found [9]. Beside the unfavourable scaling with the toroidal field, also a favourable scaling with increasing q at the edge is found [9] generally interpreted as related to the increased magnetic shear. This scaling will play unfavourably for a low q, high magnetic field tokamak, proposed here. These unfavourable features can be compensated by the higher density and hopefully by the resilience of a low q plasma to strongly unstable MHD modes.

**Discussion and Conclusions:**

First it should be emphasized that the material contained in this paper was presented at the MHD stability and control workshop in San Diego in 2011 and therefore does not pretend to be exhaustive and/or in any way complete. The aim is mainly to propose a new region of operation for tokamaks with a somehow speculative approach which is however (and this point need to be stressed again) based on some recent experimental results. In particular it has not been attempted here a detailed study of the transport, for example by employing 1 dimensional transport models. However, if we concentrate our attention to the compact high toroidal field proposal that emerge from the above considerations, this limitation is overcome by the fact that many transport studies have been already done for Ignitor [10] and our proposal does not differ from Ignitor in this respect, apart obviously the larger ohmic power available at low q.

To summarize :

In this paper a tokamak fusion concept operating at $q_a$=2 or lower is discussed. This proposal stems from recent experimental results obtained in the RFX-mod machine [1], where a well sustained $q_a$=2 at relatively high density was obtained by applying an active control on the 2/1 mode. As discussed in the Introduction, serious drawbacks exist which make difficult to scale the standard tokamak design to a reactor relevant regime. This stimulated us in exploring the “naïve” alternative of a relatively low q tokamak device.

By making very simple considerations and basing mainly our conclusions on a well established experimental database (which is also perfectly followed by the RFX-mod data), i.e. the so called Hugill diagram for density and the L mode scaling for energy confinement, we found that a medium size tokamak at high current and high magnetic field could be a suitable path to the reactor.

This conclusion, reached starting from slightly different initial assumptions, is therefore (high field and high density) similar to the Ignitor proposal [8]. In our device, as a consequence of the low q operation, for the same toroidal magnetic field the total plasma current can be 40% above the Ignitor value, enhancing the available ohmic input power and maybe opening the possibility for a fully ohmic operating device.

Another important difference with Ignitor is that the reactor proposed here will have, as a necessary component a sophisticated control for the 2/1 mode. We envisage that at least 4 active coils are necessary in poloidal directions and probably a number of the order of, at least, 4 or 6 in toroidal direction. According to the RFX-mod results also a thin resistive shell should be present near the plasma boundary. A metal liner of time constant (for magnetic field penetration) in the range of 5-10 ms will be likely enough , with the active coils being placed just outside it. Together with the active coils a suitable set of magnetic sensors should be installed to allow for an optimal control [1,2]. Obviously beside the 2/1, a good vertical control of the equilibrium is necessary as in all elongated tokamaks.

The operation at low q and the absence of a divertor in our proposal can help in avoiding all the difficulties discussed in the Introduction of this paper, although this should be at the end demonstrated experimentally. In particular the operation in L mode will certainly eliminate the ELM problem. As regard disruptions, hopefully also the operation in L mode and the low q can help in avoiding strongly unstable modes, once that the 2/1 is tamed by the control system.

It remains a very important and open question how to provide the plasma with the necessary heating power ( if the ohmic input will prove not to be enough): ICRH (Ion Cyclotron Heating) could be a realistic option, especially taking into account the fact that in this machine no metal is facing the plasma.

It should be mentioned that for a fusion device operating with tritium a carbon wall will not be a reasonable choice due to the problem of tritium retention. However a beryllium or lithium wall will probably work as well as carbon. This is an issue that should certainly be preliminary addressed. The low power density deposition at the wall, due to the absence of divertors, will however certainly help for any alternative wall material.

The formation phase of a device like that discussed here could be quite critical. We expect that a possible scenario will be to achieve very early in time and quite quickly the transition to the q(a)=2 state, and rising afterwards the toroidal field and the current at the same rate in order to maintain fixed q(a).

Another open issue is the steady state operation of such device due to the necessity of ohmic sustainment and the low bootstrap current fraction (at low q).

As a final remark it should be noted that during the long time of experiments in tokamaks, in the last 30 years, operation near the q=2 limit has been achieved in several devices (see for example [11,12,13] ) , also by using active control [14]. However this operation appeared to be unreliable and in general quite difficult, so no real systematic studies have been carried on. On the contrary the very reliable operation seen in the RFX-mod device, which is uniquely equipped with a sophisticated control system, encourage further experiments. One of the most important open issues though being the successful operation of a low q tokamak at reasonably high β's values at least around the Troyon limit. Also under this respect the RFX-mod results seem encouraging since the m=1 sawtooth activity appear similar at low q if compared with the standard q(a) operation and no sign of dangerous m=1 destabilization have been detected.

Finally we should mention that a costs analysis would be far beyond the purpose of this paper, however, it seems that the possibility of constructing an interesting fusion device with a relatively compact size (R around 2-3 m) and a magnetic field below 10 T, even if equipped with a relatively sophisticated control coil system, should keep the cost within reasonable upper limits.